\documentclass[reprint,showpacs,preprintnumbers,amsmath,amssymb,prb]{revtex4-1}

\usepackage{graphicx}
\usepackage{color}

\begin{document}

\title{
Magnetic and transport properties of
the spin-state disordered oxide
La$_{0.8}$Sr$_{0.2}$Co$_{1-x}$Rh$_x$O$_{3-\delta}$
}

\author{
Soichiro Shibasaki
}

\author{
Ichiro Terasaki
}

\email[Email me at:]{
terra@cc.nagoya-u.ac.jp
} 
\altaffiliation[Present address:]{
Department of Physics, Nagoya University, 
Nagoya 464-8602, Japan
}

\affiliation{
Department of Applied Physics, Waseda University, 
Tokyo 169-8555, Japan
}

\author{
 Eiji Nishibori
}

\author{
Hiroshi Sawa
}

\affiliation{
Department of Applied Physics, Nagoya University, 
Nagoya 464-8603, Japan
}

\author{
Jenni Lybeck
}

\author{
Hisao Yamauchi
}

\author{
Maarit Karppinen 
}

\affiliation{
Department of Chemistry, Aalto University School of Science and
Technology, FI-00076 Aalto, Finland
}

\begin{abstract}
 We report measurements and analysis of magnetization, 
 resistivity and thermopower of polycrystalline samples of 
 the perovskite-type Co/Rh oxide 
 La$_{0.8}$Sr$_{0.2}$Co$_{1-x}$Rh$_x$O$_{3-\delta}$.
 This system constitutes a solid solution for a full range of $x$,
 in which the crystal structure changes from rhombohedral 
 to orthorhombic symmetry with increasing Rh content $x$.
 The magnetization data reveal that the magnetic ground state 
 immediately changes upon Rh substitution
 from ferromagnetic to paramagnetic with increasing $x$ near 0.25,
 which is close to the structural phase boundary.
 We find that
 one substituted Rh ion diminishes the saturation moment by 9 $\mu_B$, 
 which implies that one Rh$^{3+}$ ion makes
 a few magnetic Co$^{3+}$ ions nonmagnetic (the low-spin state), 
 and causes disorder in the spin state and the highest occupied orbital.
 In this disordered composition ($0.05\le x \le 0.75$), we find that  
 the thermopower is  anomalously enhanced below 50 K.
 In particular, the thermopower of $x$=0.5 is larger
 by a factor of 10 than those of $x$=0 and 1, and
 the temperature coefficient reaches 4 $\mu$V/K$^2$
 which is as large as that of heavy-fermion
 materials such as CeRu$_2$Si$_2$.
\end{abstract}

\pacs{
75.30.Cr, 
75.30.Wx, 
72.15.Jf 
}

\maketitle

\section{Introduction}
Cobalt oxides are interesting materials.
First, the trivalent cobalt ion in perovskite-related oxides
takes different spin states depending on temperature,\cite{raccah1967,asai1998} 
pressure,\cite{vogt2003,baier2005} structure,\cite{yoshida2009} 
and magnetic field.\cite{kimura2008}
Second, hole-doping drives a spin-state crossover to make
Co$^{3+}$ ions adopt the intermediate or high-spin state,
and causes various magnetic phases.\cite{jonker1953,bhide1975,itoh1994,%
yamaguchi1996,tokura1998,masuda2003,kriener2004,berggold2005}
A doped Co$^{4+}$ ion dresses a cloud of magnetic Co$^{3+}$ ions,
as observed by neutron scattering.\cite{podlesnyak2008} 
The double exchange mechanism between the low-spin Co$^{4+}$ (${t_{2g}}^5$)
and the intermediate-spin Co$^{3+}$ (${e_g}^1{t_{2g}}^5$)
stabilizes an itinerant ferromagnetic state 
in La$_{1-x}$Sr$_{x}$CoO$_3$,
and causes spin glass or cluster glass behavior depending on 
the Sr concentration.\cite{itoh1994}
Third, since the discovery of good thermoelectric properties
in the layered cobalt oxide Na$_x$CoO$_2$,\cite{terasaki1997}
various cobalt oxides have been extensively investigated 
as possible candidates for oxide
thermoelectric materials.\cite{funahashi2000,masset2000,miyazaki2000,%
itoh2000,maignan2002,hebert2007}
The origin of the large thermopower in the layered cobalt oxides 
is discussed from viewpoints of band calculation
\cite{singh2000,kuroki2007} and spin and orbital degeneracy. \cite{koshibae2000}

Although Rh is located just below Co in the periodic table,
rhodium oxides have been less investigated than their cobalt-oxide 
counterparts, but recently have attracted considerable attention 
as candidates for thermoelectric 
oxides.\cite{okada2005,okada2005jpsj,klein2006,%
okamoto2006,shibasaki2006,kobayashi2007,maignan2009,shibasaki2010} 
An advantage for the thermoelectric application is 
that the low-spin state is highly stable in Rh$^{3+}$, 
which gives a large thermopower in the perovskite rhodium oxides.
LaCoO$_3$-based oxides show fairly good thermoelectric
properties at room temperature,\cite{androulakis2003,androulakis2004}
but the thermopower suddenly decreases to a few $\mu$V/K above 500 K,
accompanied by a spin-state crossover.\cite{{robert2007,robert2008,iwasaki2008}}
In contrast, LaRhO$_3$-based oxides are found to exhibit 
a large thermopower up to 800~K,\cite{shibasaki2009,terasaki2010}
which is theoretically explained by Usui \textit{et al}.\cite{usui2009}

Comparing the transport properties between 
La$_{1-x}$Sr$_x$CoO$_3$ and La$_{1-x}$Sr$_x$RhO$_3$,
\cite{berggold2005,iwasaki2008,shibasaki2009,terasaki2010,nakamura1993} 
we find three differences as follows;
(i) For the same doping level, the resistivity of doped LaCoO$_3$ 
is one order of magnitude smaller than that of doped LaRhO$_3$.
(ii) The metal-insulator transition occurs around $x_c=0.3$ 
in La$_{1-x}$Sr$_x$CoO$_3$; in La$_{1-x}$Sr$_x$RhO$_3$,
$x_c$ is significantly smaller ($\sim$0.15). 
(iii) La$_{1-x}$Sr$_x$CoO$_3$ is a ferromagnetic metal, while
La$_{1-x}$Sr$_x$RhO$_3$ is a Curie-Weiss metal.
These three differences are basically ascribed 
to the difference in the highest occupied orbital of Co/Rh ions. 
In the doped LaCoO$_3$, 
holes in the $e_g$ orbital are responsible for the conduction,
and interact with the Co$^{3+}$ ions through the double exchange mechanism.
In the doped LaRhO$_3$, in contrast, holes in the $t_{2g}$ orbital
are responsible for the conduction.
It is thus tempting to study the physical properties 
of a solid solution of doped LaCoO$_3$ and LaRhO$_3$,
where the highest-occupied orbitals and the spin states 
are disordered.
Very recently, Li \textit{et al}.\cite{li2010} reported the structural and
thermoelectric properties of LaCo$_{1-x}$Rh$_x$O$_3$.
We expect that more exotic effects will emerge in the solid-solution
system of Co and Rh, when carriers are doped.
In this paper, we show the magnetic and transport properties 
of polycrystalline samples of La$_{0.8}$Sr$_{0.2}$Co$_{1-x}$Rh$_x$O$_{3-\delta}$.

\section{Experimental}
Polycrystalline samples of
La$_{0.8}$Sr$_{0.2}$Co$_{1-x}$Rh$_x$O$_{3-\delta}$ 
were prepared by a solid-state reaction.
The Sr content of 0.2 was chosen based on the fact that
the solubility limit of Sr in La$_{1-y}$Sr$_y$RhO$_3$
is at $y$=0.2.\cite{terasaki2010}
At this doping level, La$_{0.8}$Sr$_{0.2}$CoO$_3$ is a ferromagnetic semiconductor,
and La$_{0.8}$Sr$_{0.2}$RhO$_3$ is a Curie-Weiss metal.
Thus we expect to observe a systematic evolution between the two
states with increasing Rh content $x$. 
Stoichiometric amounts of La$_2$O$_3$, SrCO$_3$, 
Co$_3$O$_4$ and Rh$_2$O$_3$ were mixed and 
calcined at 1373~K for 24~h in air.
The calcined products were thoroughly ground, 
pelletized and sintered at 1473~K for 48~h in air.

X-ray diffraction patterns 
were measured using MoK$\alpha$ radiation
(RIGAKU ultraX18 with multilayer mirror)
and a Debye-Scherrer camera of radius 286.48 mm with
an imaging plate.
Oxygen contents of the samples were determined 
by thermogravimetric H$_2$-reduction analysis 
with reproducibility better 
than $\pm$0.03 for $3-\delta$.\cite{karppinen2002} 
The resistivity was measured using a conventional four-probe technique
in a liquid He cryostat from 4.2 to 300~K.
The thermopower was measured using a steady-state technique 
with a typical temperature gradient of 0.5-1~K/cm
in a liquid He cryostat from 4.2 to 300~K.
Using Magnetic Property Measurement System (MPMS, Quantum Design),
the magnetization measurements were performed 
with an external field of 0.1~T from 5 to 400 K, 
and the magnetization-field curve was 
measured from 0 to 7~T at 5~K.

\section{Results and discussion}
From the thermogravimetric reduction experiments, it was revealed 
that the La$_{0.8}$Sr$_{0.2}$Co$_{1-x}$Rh$_x$O$_{3-\delta}$ system 
is apparently oxygen-deficient. 
This is rather common for ambient-pressure-synthesized ACoO$_3$-type 
perovskites based on A-site cations 
with a lower than $+3$ valence state in average. 
For the present samples, the $3-\delta$ value was determined at 2.95(4), 
with no evident dependency on $x$. 
Consequently we may conclude that 
in La$_{0.8}$Sr$_{0.2}$Co$_{1-x}$Rh$_x$O$_{3-\delta}$
the average valence of the B-site cations (Co and Rh) 
remains essentially constant 
at ca. $+3.1$ through the whole substitution range, 
being significantly lower than the value ($+3.2$) calculated assuming
$\delta$=0.

\begin{figure}
\begin{center}
 \includegraphics[width=8cm,clip]{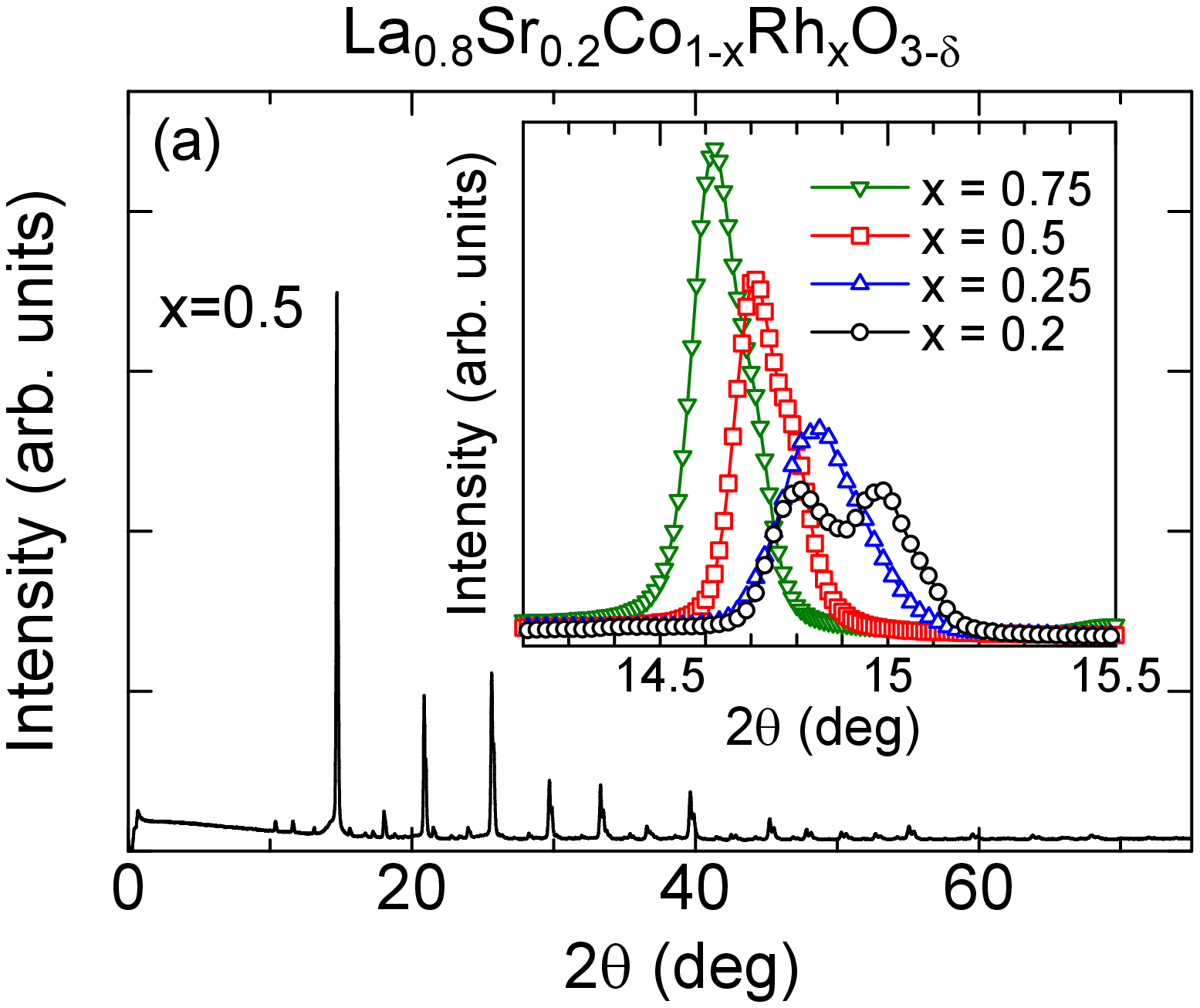}
 \includegraphics[width=8cm,clip]{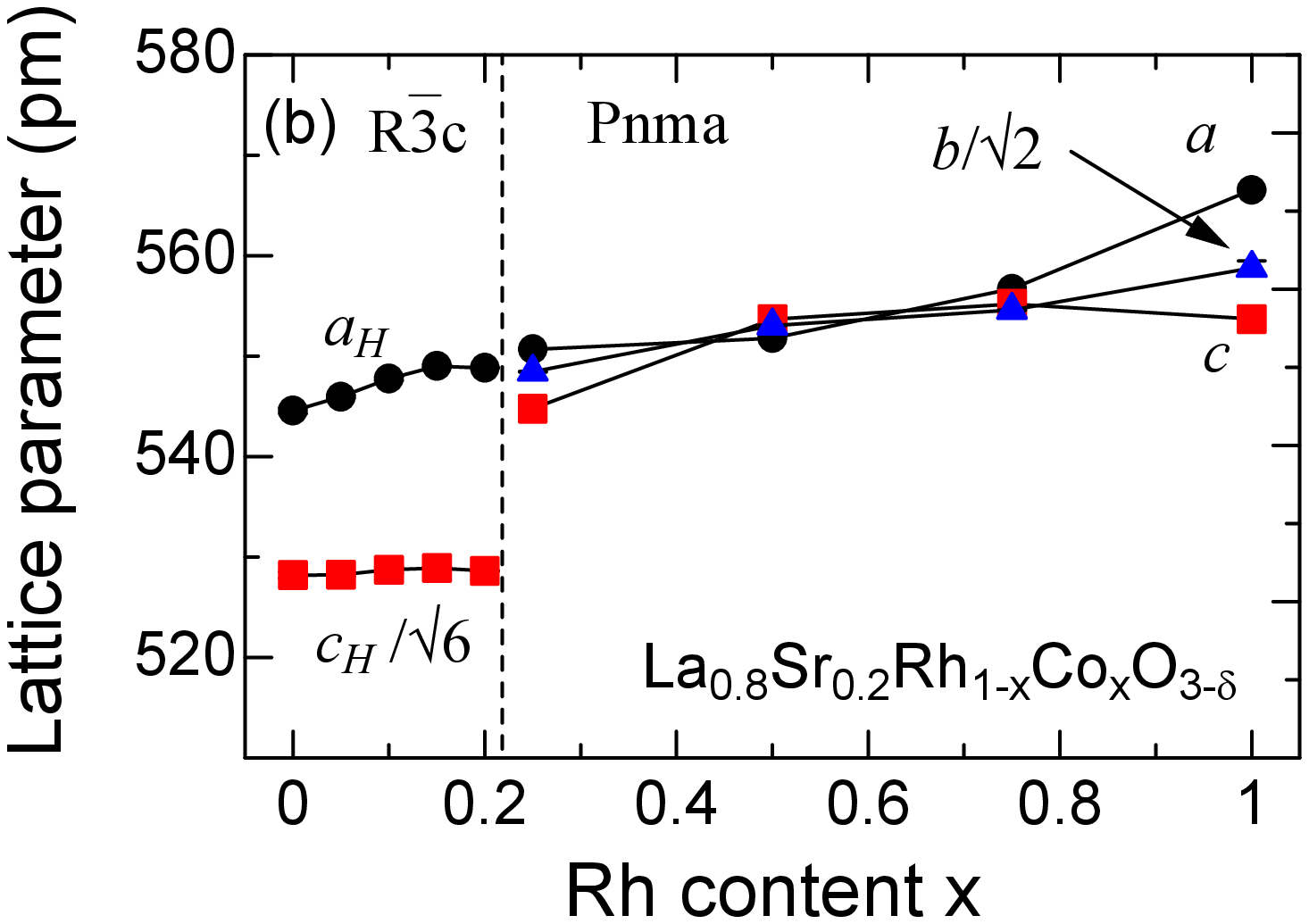}
\end{center}
\caption{(Color online) (a) A typical x-ray diffraction 
pattern and (b) the lattice
parameters of La$_{0.8}$Sr$_{0.2}$Co$_{1-x}$Rh$_x$O$_{3-\delta}$.
The crystal structure is rhombohedral for $x\le 0.2$
and orthorhombic for $x\ge 0.25$.
The inset shows the $x$ dependence of the main peak 
near $2\theta=$15 deg.
}
\label{fig1}
\end{figure}

Figure~\ref{fig1}(a) shows a typical x-ray diffraction pattern 
for $x$=0.5.
At room temperature, La$_{0.8}$Sr$_{0.2}$CoO$_{3-\delta}$ 
crystallizes in rhombohedral  (space group $R\bar{3}c$), 
La$_{0.8}$Sr$_{0.2}$RhO$_{3-\delta}$ 
in orthorhombic (space group $Pnma$) structure,
as is revealed from the evolution of 
the main peak with $x$ in the inset of Fig.~\ref{fig1}(a).
Clearly the phase boundary is between $x$=0.2 and 0.25 at room temperature.
The lattice parameters of these materials are shown 
in Fig.~\ref{fig1}(b).
The lattice parameters increase with increasing Rh content, 
which is consistent with the larger ionic radius of 
Rh$^{3+}$ compared to that of Co$^{3+}$.
This means that the substitution works well, and 
the Co and Rh species are distributed homogeneously to form a solid solution.

\begin{figure}[t]
\begin{center}
 \includegraphics[width=8cm,clip]{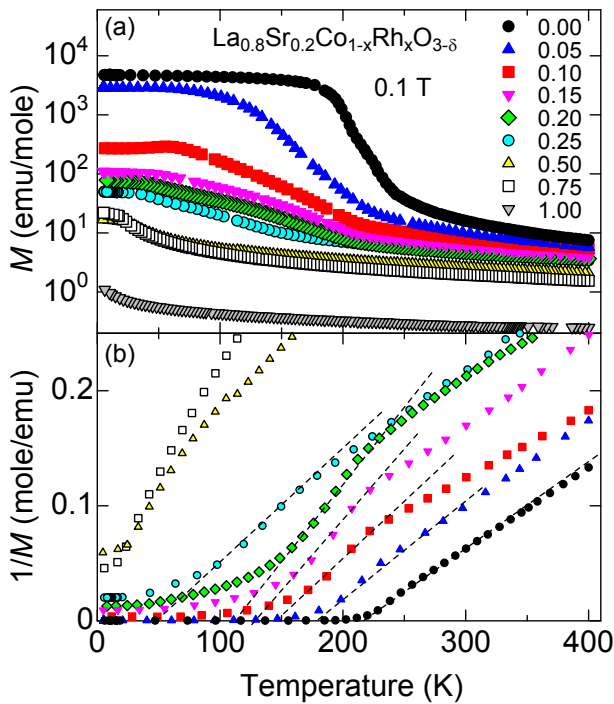}
\end{center}
\caption{(Color online) (a) Magnetization $M$ and 
(b) inverse magnetization $1/M$ of 
La$_{0.8}$Sr$_{0.2}$Co$_{1-x}$Rh$_x$O$_{3-\delta}$.
The dotted lines are guide to the eyes, and indicates
the Curie temperature at $1/M$=0.
}
\label{fig2}
\end{figure}

Figure \ref{fig2} shows the temperature dependence of the 
magnetization of La$_{0.8}$Sr$_{0.2}$Co$_{1-x}$Rh$_x$O$_{3-\delta}$.
A ferromagnetic transition is observed for La$_{0.8}$Sr$_{0.2}$CoO$_{3-\delta}$ 
at around 220~K, as is already reported in literature \cite{kriener2004}.
The transition temperature and the saturation moment
rapidly decrease with increasing Rh content.
This means that the substituted Rh ions seriously disrupt 
the magnetic order.
The inverse magnetization is plotted as a function 
of temperature in Fig.~\ref{fig2}(b).
The Curie temperature ($T_{\rm C}$) is estimated by the extrapolation of the high
temperature data as shown by the dotted lines.
$T_{\rm C}$ disappears near $x$=0.25, which coincides
with the structural phase boundary shown in Fig. 1(b).
The inverse magnetization shows finite values for $0.1\le x\le 0.25$,
which indicates that the ferromagnetism within this composition range
is not homogeneous, and only a part of the volume goes ferromagnetic.
On the contrary, the inverse magnetization for $x$=0.5 and 0.75
has a positive offset, indicating an antiferromagnetic interaction.
The magnetization of $x$=1 shows a value one order of magnitude smaller
than that of $x$=0.75, which is consistent with the fact that
the Rh$^{3+}$ ions are stable in the low-spin state.
However, the magnetization increases with decreasing temperature.
Nakamura \textit{et al}. \cite{nakamura1993} first measured the susceptibility of
this compound, and analyzed the data with a sum of 
a Curie-Weiss term and a constant term. 
This suggests that the Rh$^{4+}$ ion does not act as a simple itinerant hole,
but works partially as a localized moment.
Such a duality of itinerancy and localization is observed
in the misfit-layered cobalt oxides.\cite{bobroff2007}
 
To discuss the spin states of the present compound,
we will briefly review the discussion on the spin 
states of undoped and doped LaCoO$_3$ here.
As is well known, 
the spin states of the Co$^{3+}$ ion have been debated 
for more than forty years,
because the low and high-spin states of Co$^{3+}$ 
are almost degenerate in energy.
First of all, there is a consensus that the Co$^{3+}$ 
ion in LaCoO$_3$ takes the low-spin state below 10 K, 
which was suggested by early susceptibility measurement\cite{jonker1966}
and was verified by neutron experiment.\cite{asai1989}
Above around 90 K, LaCoO$_3$ becomes magnetic,
which implies that the high-spin or intermediate-spin state
is thermally excited above that temperature.
Goodenough \cite{goodenough1958} originally thought 
that the high-spin state of Co$^{3+}$ was thermally created.
Later Potze et al. \cite{potze1995}
suggested a possibility of the intermediate
spin state by introducing the hybridization to the oxygen $2p$ state.
Asai et al. \cite{asai1994} reported that there were 
two magnetic transitions, one near 90 K and a second near 500 K.
They further proposed that the lower transition is a 
crossover from low-spin to the intermediate-spin state.\cite{asai1998}

The concept of the intermediate-spin state is highly controversial.
On the basis of a localized picture,
the intermediate-spin state has much higher energy than
the low-spin  and high-spin states .\cite{haverkort2006}
It is convenient to introduce a concept of `$p-$hole'
in the presence of hybridization between the 
transition-metal $3d$ and O $2p$ states.\cite{zaanen1985}
Using this concept, the intermediate-spin state of Co$^{3+}$
is expressed as a composite of
the high-spin state of Co$^{2+}$ (${e_g}^2{t_{2g}}^5$)
antiferromagnetically coupled with one $p-$hole.
Korotin et al. \cite{korotin1996}
pointed out that the one $e_g$ electron
on the intermediate-spin state of Co$^{3+}$ has 
an orbital degrees of freedom, and suggested a possibility
of orbital ordering as an origin of the nonmetallic nature of LaCoO$_3$.
Maris et al. \cite{maris2003} observed a monoclinic distortion 
in LaCoO$_3$, which they attributed to a piece of evidence for the
orbital ordering.
Very recently, Nakao et al. \cite{nakao2011}
observed $\sigma \to \pi'$ resonant x-ray
scattering in the related cobalt oxide Sr$_3$YCo$_4$O$_{10.5}$,
which is clear evidence of the orbital ordering of $e_g$ electrons
and the existence of the intermediate-spin state as well.
On the contrary, some researchers questioned the existence of the
intermediate-spin state.
Haverkort et al. \cite{haverkort2006} clarified that the spin state of LaCoO$_3$
is well understood by a mixture of the low-spin and high-spin states
from the Co-$L_{2,3}$ x-ray absorption spectra.
They further pointed out that the spectra carried a large orbital
moment, which is seriously incompatible with the intermediate-spin state.
A similar orbital contribution was detected by electron spin resonance (ESR).
From high-field ESR measurement, Noguchi et al.\cite{nojiri2002} 
found the $g$-factor to be 3.35 for LaCoO$_3$ at 50 K. 
This value is substantially larger than 2,  
strongly suggesting the existence of the orbital moment ($L=1$),
and equivalently, the existence of the high-spin state.

The second transition near 500 K is also controversial.
At this temperature, LaCoO$_3$ experiences a metal-insulator transition,
and the metallic nature at high temperature is not well understood at present.
Heikes et al. \cite{heikes1964} claimed that 
pair formation of Co$^{2+}$ an Co$^{4+}$ should be taken into
account for the high-temperature metallic conduction,
but this idea was rejected by Jonker\cite{jonker1966}
Raccah and Goodenough \cite{raccah1967}
proposed a spin-state order, above which
the high-spin and low-spin state Co$^{3+}$ ions
were ordered in the NaCl-type pattern.
However, this ordering was not detected in the neutron 
experiments.\cite{thornton1986}
Asai et al. \cite{asai1998} associated this transition with a crossover
from low temperature intermediate-spin to high-temperature high-spin state. 
Tokura et al. \cite{tokura1998} proposed that this metal-insulator transition
can be regarded as a Mott transition
where one $e_g$ electron per Co in the intermediate-spin state
is identical to a half-filled state.
Kobayashi et al. \cite{kobayashi2006} found a similarity 
in the high-temperature metallic state between 
LaCoO$_3$ and Sr$_3$YCo$_4$O$_{10.5}$.
In the latter compound, low metallic resistivity
is realized above 600 K, together with a small
thermopower and Hall coefficient.
This indicates that most of the Co$^{3+}$ ions become ``itinerant'' to
make a large Fermi surface in the charge sector.
We should note that this metallic state is quite anomalous 
in the sense that it accompanies a large effective moment
of 3.4 $\mu_B$ per Co in the spin sector.

Compared with the diverging discussion on LaCoO$_3$,
the situation is better for 
the spin state of the doped LaCoO$_3$.
In this case, the doped hole hops from one Co to another
through the hybridization to the oxygen 2$p$ state,
and the intermediate-spin state can be stabilized
relative to the high-spin state.\cite{podlesnyak2008}
It is now established that the ground state
changes from non-magnetic insulator for LaCoO$_3$ to ferromagnetic
metal for La$_{1-x}$Sr$_{x}$CoO$_3$ ($x> 0.25$).\cite{jonker1953,bhide1975,itoh1994,%
yamaguchi1996,tokura1998,masuda2003,kriener2004,berggold2005}
Since the magnetic and transport properties of
the ferromagnetic La$_{1-x}$Sr$_{x}$CoO$_3$ are similar to those
of La$_{1-x}$Sr$_{x}$MnO$_3$, 
it is natural to associate them with the double exchange mechanism.\cite{tokura1994} 
For this mechanism,
a hole in the $e_g$ band is responsible
between the intermediate-spin state of Co$^{3+}$ (${e_g}^1{t_{2g}}^5$)
and the low-spin state of Co$^{4+}$ (${e_g}^0{t_{2g}}^5$), 
whereas a hole in the $t_{2g}$ band is responsible
between the high-spin state of Co$^{3+}$
(${e_g}^2{t_{2g}}^4$) and the low-spin state of Co$^{4+}$ (${t_{2g}}^5$).
Since the reduction of the kinetic energy of the hole is
essential to the double exchange mechanism, 
the former case is more likely to be realized, because
the $e_g$ band is wider than the $t_{2g}$ band.
Kriener et al. \cite{kriener2004} showed that the saturation magnetization
for ferromagnetic La$_{0.75}M_{0.25}$CoO$_3$ ($M$=Sr and Ba)
was 1.65 $\mu_B$,
which is consistent with the intermediate-spin state of Co$^{3+}$
the low-spin state of Co$^{4+}$.
Podlesnyak et al.\cite{podlesnyak2008} revealed 
that  a spin cluster of one low-spin state Co$^{4+}$ ion
surrounded with six intermediate-spin state Co$^{3+}$ ions
was localized at low temperature in La$_{0.998}$Sr$_{0.002}$CoO$_3$.
However, things are not simple; Wu and Leighton \cite{wu2003} 
claimed that the glassy ferromagnetism of La$_{1-x}$Sr$_x$CoO$_3$ 
is understood by a combination of the intermediate-spin state of Co$^{3+}$
the {\it intermediate-spin state} of Co$^{4+}$.
Kriener et al. \cite{kriener2004} also pointed out 
that the saturation magnetization
for  La$_{0.75}$Ca$_{0.25}$CoO$_3$ is 1 $\mu_B$,
which is significantly smaller than the Sr-doped sample. 
As mentioned above, in spite of some controversy, 
many experiments suggest that Co$^{3+}$ in the doped LaCoO$_3$
is in the intermediate-spin state.
Thus we will assume that the Co$^{3+}$ ions are in the
intermediate-spin state in La$_{0.8}$Sr$_{0.2}$Co$_{1-x}$Rh$_x$O$_3$.

\begin{figure}
\begin{center}
 \includegraphics[width=7cm,clip]{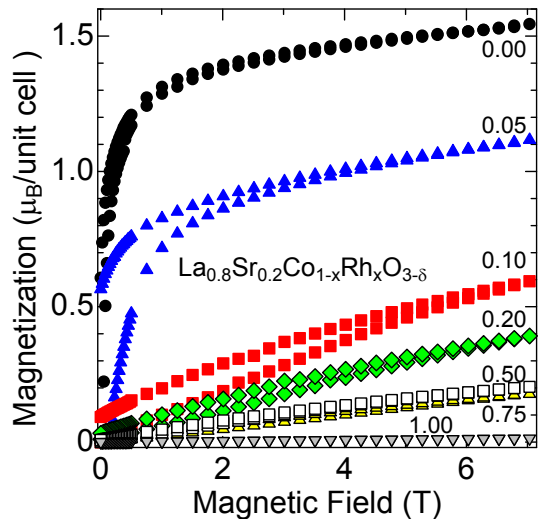}
\end{center}
\caption{(Color online) Magnetization-field curves 
for La$_{0.8}$Sr$_{0.2}$Co$_{1-x}$Rh$_x$O$_{3-\delta}$ at 5~K.}
\label{fig3}
\end{figure}

Figure~\ref{fig3} shows the magnetization-field ($M-H$) curves
of La$_{0.8}$Sr$_{0.2}$Co$_{1-x}$Rh$_x$O$_{3-\delta}$ at 5~K.
The magnetic moment of La$_{0.8}$Sr$_{0.2}$CoO$_{3-\delta}$ at 7~T 
is 1.55~$\mu_{\rm B}$, which is well understood by
the combination of the intermediate-spin state for Co$^{3+}$ 
($S=1$) and the low-spin state of Co$^{4+}$ ($S=1/2$).
For $x=$0.05, the hysteresis loop broadens possibly because
the substituted Rh acts as a strong pinning center.
Looking at the magnetization at 7 T, we find that
the saturation magnetization decreases by 0.45 $\mu_B$ from $x$=0
to 0.05, which implies that one Rh ion 
decreases the magnetic moment by 9 $\mu_B$.
Assuming the intermediate-spin state for Co$^{3+}$ (2 $\mu_B$), 
we estimate that three to four Co$^{3+}$ ions transfer
to the low-spin state per one substituted Rh ion.
For $x$ =0.1, the $M-H$ curves are almost linear
with a tiny hysteresis, suggesting a tiny fraction of
the ferromagnetic order.
From $x$=0.05 to 0.1, the saturation magnetization
decreases by 0.5 $\mu_B$, which is close to the rate from $x$=0 to 0.05.
As a result, the spin state of the Co$^{3+}$ ions
is disordered in La$_{0.8}$Sr$_{0.2}$Co$_{1-x}$Rh$_x$O$_{3-\delta}$,
as the intermediate-spin Co$^{3+}$
is partially converted to the low-spin state by the substituted Rh ions.
Concomitantly, disorder is induced in the highest-occupied orbitals
of $t_{2g}$ and $e_g$.
We notice that the magnetization does not saturate,
but continues to increase above 4 T
with almost the same slope from $x$=0 to 0.1, although
we do not understand the origin for this.
For $x>$0.1, the magnetization is almost linear in external field, 
the slope decreasing with $x$.

\begin{figure}
\begin{center}
 \includegraphics[width=8cm,clip]{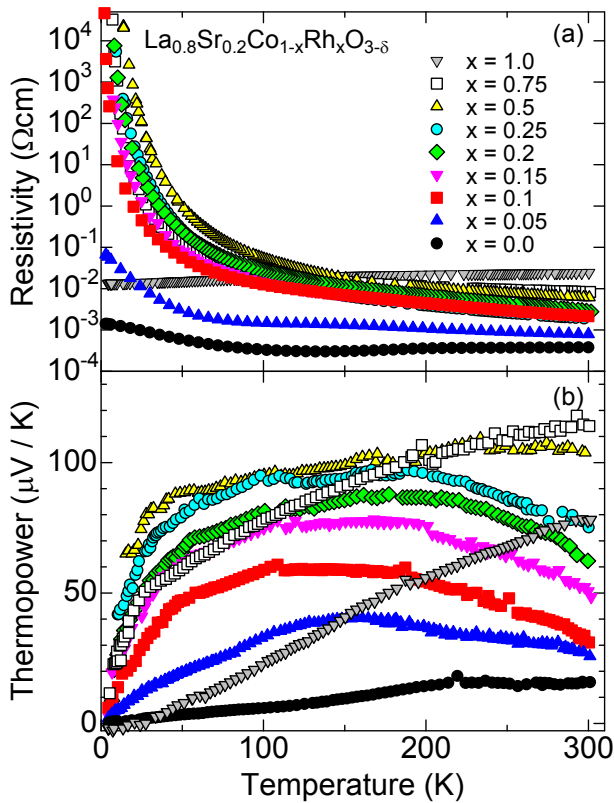}
\end{center}
\caption{(Color online) (a) Resistivity and 
(b) thermopower for La$_{0.8}$Sr$_{0.2}$Co$_{1-x}$Rh$_x$O$_{3-\delta}$.}
\label{fig4}
\end{figure}

Next let us focus on the transport properties.
Figure~\ref{fig4}(a) shows the temperature dependence 
of the resistivity of La$_{0.8}$Sr$_{0.2}$Co$_{1-x}$Rh$_x$O$_{3-\delta}$.
The values for $x$=0 and 1 are consistent with 
those given in literature,\cite{iwasaki2008,terasaki2010}
and the values for the intermediate compositions 
increase monotonically with increasing Rh content $x$ at 300~K.
This indicates that the room-temperature resistivity is roughly
understood as an average of the two end phases.
We should note that the resistivity of the $x$=0 sample
exhibits an upturn below 100 K, which is ascribed to the 
electronic phase separation of the spin clusters.\cite{wu2003,wu2006}
The spin cluster consisting of the low-spin Co$^{4+}$ surrounded 
with the intermediate-spin Co$^{3+}$ cloud can move rather freely at room
temperature, and is gradually confined by the low-spin Co$^{3+}$ ions
in the background with decreasing temperature.
Thus, although the room-temperature resistivity is 
less than 1 m$\Omega$cm, the carriers are eventually localized at low
temperatures.
In contrast, the resistivity for $x$=1 shows a metallic conduction down
to 4 K with a large value of the order of 10 m$\Omega$cm, owing to 
the narrow $t_{2g}$ band.
The effect of the solid solution is drastic at low temperatures.
The $x$=0.05 sample shows a strong localization below 50 K,
which suggests that the substituted Rh works as a strong pinning center.
As was discussed above, one substitued Rh creates 
a few nonmagnetic Co$^{3+}$ ions at 5 K, 
which can disrupt the motion of the spin cluster.
For $x>$0.10, the resistivity shows non-metallic temperature
dependence at all temperatures, which is explained by the 
variable-range-hopping transport.
This implies that the conduction paths are seriously disordered
owing to the disorder in the highest occupied orbital.
A similar resistivity is seen in 
the layered oxide Bi-Ba-(Co,Rh)-O.\cite{okada2009}

Figure~\ref{fig4}(b) shows the temperature dependence 
of the thermopower of La$_{0.8}$Sr$_{0.2}$Co$_{1-x}$Rh$_x$O$_{3-\delta}$,
in which the data for the two end phases
are consistent with those given in literature.\cite{iwasaki2008,terasaki2010}
The thermopower for $x$=0 is small and roughly linear in temperature,
indicating a metallic nature of this composition.
A small cusp is seen near 220 K, 
which is close to the Curie temperature.
On the other hand, the data for $x$=1 show a large value of 80 $\mu$V/K
at room temperature, which suggests that the band effective mass is
substantially large owing to the narrow $t_{2g}$ bands.
The temperature dependence is roughly linear in temperature,
indicating that the electronic states are essentially understood 
as a metal.
The effects of the solid solution to the thermopower are remarkable.
The thermopower is enhanced particularly at low temperatures.
For example, the thermopower for $x$=0.05 at 50 K is \textit{larger} than
those for $x$=0 and 1, which strongly suggests that
the electronic states for $0.05 \le x \le 0.75$
cannot be understood from a simple average of $x$=0 and 1.
With increasing $x$ from 0.05, the thermopower systematically increases
up to 0.5, and exceeds $k_B/e$=86 $\mu$V/K at 50 K for $x$=0.5, 
which is ten times larger than those of $x$=0 and 1.
We should note that simple disorder cannot enhance
the thermopower because this quantity is insensitive 
to the grain boundary scattering.\cite{carrington1994}
We should also note that the oxygen vacancies cannot be
the origin of this enhancement either, 
as our oxygen-content analysis indicated that oxygen content
and therefore the carrier density remains essentially constant
in La$_{0.8}$Sr$_{0.2}$Co$_{1-x}$Rh$_{x}$O$_{3-\delta}$.
A similar phenomenon is seen in the misfit-layered Rh/Co oxides,
but the effect of the solid solution is subtle;
the thermopower for $x$=0.5 slightly exceeds 
those for $x$=0 and 1 near 300 K.\cite{okada2009}

The temperature dependence of the thermopower
for $0.05\le x \le 0.75$ is rather complicated.
The thermopower is roughly linear in temperature below about 30 K, 
nearly independent of temperature from 50 to 200 K.
It slowly decreases with increasing temperature above 200 K
for $0.05\le x\le 0.25$.
The thermopower of La$_{1-x}$Sr$_x$CoO$_3$
decreases with increasing temperature above 200 K,\cite{iwasaki2008} 
which is attributed to a precursor of the high-temperature
metallic state above 500 K.
Since Co is the main B-site constituent for $x\le$0.25,
the high-temperature electronic states tend to favor those
of La$_{1-x}$Sr$_x$CoO$_3$.
For $x$=0.5 and 0.75, on the other hand, the thermopower continues to
increase up to 300 K, which resembles the thermopower of 
 La$_{1-x}$Sr$_x$RhO$_3$.\cite{terasaki2010}

\begin{figure}
\begin{center}
 \includegraphics[width=7cm,clip]{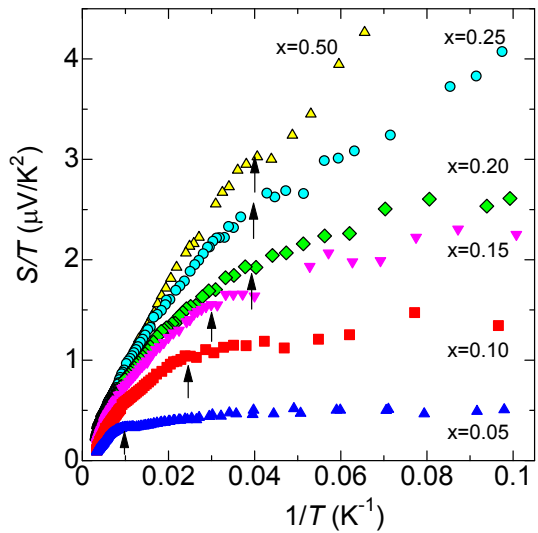}
\end{center}
\caption{(Color online) 
The thermopower divided by temperature $S/T$ 
plotted as a function of $1/T$.
The arrows represent the degenerate temperature
below which the thermopower 
is roughly linear in temperature.
}
\label{fig5}
\end{figure}

Let us have a closer look at the thermopower at low temperatures.
Below 100 K, gross features of the thermopower are
similar to those of the layered cobalt and rhodium 
oxides.\cite{masset2000,miyazaki2000,itoh2000,maignan2002,hebert2007,okada2005}
Limelette \textit{et al}. \cite{limelette2006}
have proposed that the low-temperature thermopower of 
the misfit-layered cobalt oxides can be understood 
by a sum of a constant term and a $T$-linear term,
corresponding to the spin part and the Fermi liquid part,
respectively.
Figure \ref{fig5} shows $S/T$ plotted as a function of $1/T$.
For small $1/T$ (i.e. high temperature), $S/T$ is roughly linear in $1/T$,
indicating that $S$ is weakly dependent on $T$.
With increasing $1/T$, the behavior starts 
to deviate from the linear relation as indicated by the arrows, 
and $S/T$ tends to saturate.
In other words, $S/T$ goes towards a constant value for a large
$1/T$ (low temperature).
In this context, the temperature indicated by the arrows 
corresponds to the onset temperature of the $T$-linear thermopower. 
It should be noted that $S/T$ for $x$=0.25 is anomalously large.
It seems to approach $\sim$4 $\mu$V/K$^2$ at zero temperature,
although $S/T$ still increases in the measured temperature range. 
Behnia \textit{et al}. \cite{behnia2004} have found a universal relationship
between $S/T$ and the electron specific heat coefficient $\gamma$,
such that the thermopower is given by
\begin{equation}
 S = K_0\frac{k_B}{e}\frac{T}{T_0},
\end{equation}
where $T_0$ is the degenerate temperature (the Fermi temperature)
and $K_0$ is a constant of the order of unity.
Equation (1) implies that $T_0$ is of the order of $10-10^2$ K
from $S/T$=4 $\mu$V/K$^2$.
Since the temperature indicated by the arrows in Fig. 5
is of the same order of $T_0$,
we identify this temperature to $T_0$ like
heavy fermion and/or mixed valence materials.
In fact the value of $S/T=$4 $\mu$V/K$^2$ is close to
$S/T$ of the heavy-fermion materials,
CeRu$_2$Si$_2$ and CeCoIn$_5$.\cite{behnia2004} 
However, we notice that the resistivity is highly non-metallic,
and the electronic states are even qualitatively different from
these materials.

Here we will examine Behia's relationship of $S/T$ to $\gamma$ 
for La$_{1-x}$Sr$_x$CoO$_3$.
The $S/T$ value for $x=0$ is about 0.1 $\mu$V/K$^2$, 
which corresponds to $\gamma=$10 mJ/mol K$^2$.
This value is smaller than the observed 30 mJ/mol K$^2$
reported by He et al.\cite{he2009}
This is not surprising, because the ground state of 
La$_{1-x}$Sr$_{x}$CoO$_3$ is a ferromagnetic metal, 
not a paramagnetic Fermi liquid.
A similar situation is seen in the doped LaMnO$_3$; 
the thermopower is negligibly small ($\pm 5~\mu$V/K at 77 K),\cite{mandal2000} 
and $\gamma$ is  5 mJ/mol K$^2$ for the metallic region.\cite{okuda2000}
Okuda et al. \cite{okuda2000} pointed out 
that the Kadowaki-Woods relation is seriously broken down 
in the doped LaMnO$_3$, where the $T^2$ coefficient of the
resistivity is too large to compare with other correlated metals.
This indicates that the double-exchange mediated ferromagnetic metals
cannot be understood from the conventional Fermi liquid ground state.
On the other hand, our sample of $x=0.5$ is paramagnetic, which can be
compared with some heavy fermion compounds.

\begin{figure}[t]
\begin{center}
 \includegraphics[width=7cm,clip]{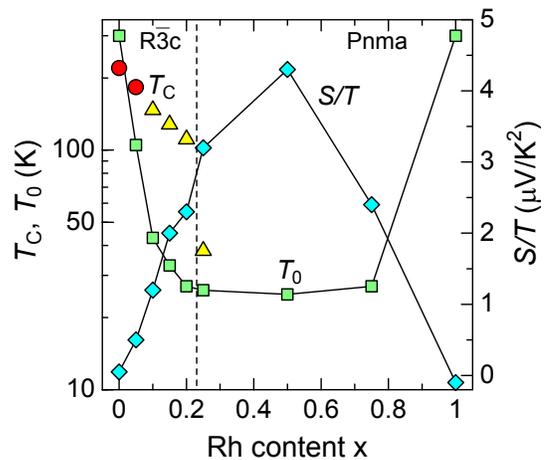}
\end{center}
\caption{(Color online)
Phase diagram of La$_{0.8}$Sr$_{0.2}$Rh$_{1-x}$Co$_x$O$_{3-\delta}$.
$T_{\rm C}$ and $T_0$ correspond to the Curie temperature
determined in Fig. 3, 
and the degenerate temperature determined in Fig. 5,
respectively.
The circles indicate 
the Curie temperature as a bulk transition, 
and the triangles represent a kink temperature
seen in Fig. 2(b).
In the right axis, the temperature coefficient 
of the thermopower $S/T$ at 15 K is also plotted.
}
\label{fig6}
\end{figure}

Figure \ref{fig6} summarizes the electric phase diagram 
of the title compound,
where $T_{\rm C}$ and $T_0$ correspond to 
the Curie temperature determined in Fig. 3 and
the degenerate temperature determined in Fig. 5,
respectively.
Although $T_0$ for $x$=0 and 1 
should be above 300 K, we plot it at 300 K.
We plot $T_{\rm C}$ with different marks,
circles and triangles.
The circle corresponds to a bulk ferromagnetic transition,
i.e., the divergence of the susceptibility in the absence of external field.
In Fig. 2(b), $1/M$ for $x=0$ and 0.05 
touches the abscissa axis below $T_{\rm C}$.
On the other hand, $1/M$ for $x=$0.1, 0.15 and 0.2
shows a kink at $T_{\rm C}$, but remains finite below.
This indicates that only a part of the sample goes
ferromagnetic, and thus we plot the kink temperature
with triangles as a Curie temperature 
for a partial ferromagnetic order.
In the same figure, we plot $S/T$ evaluated at 15 K.
As mentioned above, the ferromagnetism appears only 
in the rhombohedral phase, although the transition
is not of bulk nature for $x\ge$ 0.1.
The degenerate temperature is roughly independent
of $x$ from 0.2 to 0.75, and correlates with $S/T$.
This means that the enhancement in $S/T$ is related
to neither the ferromagnetism nor the crystal structure,
but is related to the disorder induced by 
the solid solution of Co and Rh.

Finally let us discuss a possible origin for the enhanced
$S/T$ for $0.05\le x\le0.75$.
As discussed above,
the spin state and the highest occupied orbital
are disordered in this solid solution, 
which induces additional entropy in the system.
Since the thermopower reflects
entropy per carrier,\cite{terasaki2010,terasaki2010b}
we expect that the enhanced thermopower found here 
is related to the entropy due to the disorder in the spin state 
and/or the highest occupied orbital.
In the perovskite-related oxide
Sr$_3$YCo$_4$O$_{10.5}$,\cite{yoshida2009}
partial substitution of Ca for Sr substantially enhances
the thermopower, which is associated with the spin state 
crossover of the Co$^{3+}$ ions.
In the same oxide, also partial substitution of Rh for Co 
enhances the thermopower.\cite{terasaki2010b}
These two examples indicate that the spin state distribution
of the Co$^{3+}$ ions affects the thermopower,
and we think that a similar mechanism works 
in the present system as well.
Further experimental and theoretical studies are
of course necessary to examine this conjecture.

\section{Summary}
We have presented the transport and magnetic properties 
of the perovskite-type Co/Rh oxide
La$_{0.8}$Sr$_{0.2}$Co$_{1-x}$Rh$_x$O$_{3-\delta}$.
From the magnetization-field curve, we find that 
one substituted Rh ion makes 3-4 Co$^{3+}$ ions transfer to 
the low-spin state at 5 K,
and causes disorder in the spin state.
As a result, the ferromagnetic order in $x$=0 
immediately vanishes at $x=0.10$,
while a part of the sample remains ferromagnetic near 
the crystal structure phase boundary at $x=0.25$.
While the resistivity systematically changes with $x$ at 300 K,
it is highly non-metallic for $0.05\le x\le 0.75$ at low temperatures.
This suggests that the spin state disorder acts as 
a strong pinning center.
The most remarkable effect is that
the thermopower is anomalously enhanced for $0.05\le x\le 0.75$.
In particular, the thermopower for $x$=0.5 exceeds $k_B/e$ at 50 K,
which is ten times as large as those for $x$=0 and 1.
The enhanced thermopower looks similar to the thermopower
of heavy fermion or mixed valence materials, 
but the origin of the enhancement should be different.
We suggest that disorder in the spin state or the highest-occupied orbital
is the origin for the additional entropy 
that attaches the conduction electrons to enhance the thermopower.

\section*{Acknowledgments}
The authors would like to thank Y. Klein, M. Abdel-Jawad, 
and T. Tynell for fruitful discussion.
K. Noguchi is thanked for sample preparation 
at an early stage of this study.
This work was partially supported by a Grant-in-Aid for 
Scientific Research, MEXT (No. 21340106), Japan and 
by Strategic Japanese-Finland Cooperative Program 
on ``Functional Materials'', JST, Japan,
and Academy of Finland (No. 130352). 


%

\end{document}